\documentclass[aps,twocolumn,preprintnumbers,superscriptaddress,amsmath,amssymb,showpacs]{revtex4-1}

\usepackage{graphicx}
\usepackage{subfigure}
\usepackage{epsfig}
\usepackage{dcolumn}
\usepackage{bm}
\usepackage{ulem}
\usepackage{color}
\usepackage{textcomp}
\usepackage{indentfirst}
\usepackage{epstopdf}

\begin{document}

\title{Monolayer charge-neutral graphene on platinum with extremely weak electron-phonon coupling}

\author{Wei Yao}
\affiliation{State Key Laboratory of Low Dimensional Quantum Physics and Department of Physics, Tsinghua University, Beijing 100084, China}

\author{Eryin Wang}
\affiliation{State Key Laboratory of Low Dimensional Quantum Physics and Department of Physics, Tsinghua University, Beijing 100084, China}

\author{Ke Deng}
\affiliation{State Key Laboratory of Low Dimensional Quantum Physics and Department of Physics, Tsinghua University, Beijing 100084, China}

\author{Shuzhen Yang}
\affiliation{State Key Laboratory of Low Dimensional Quantum Physics and Department of Physics, Tsinghua University, Beijing 100084, China}

\author{Wenyun Wu}
\affiliation{State Key Laboratory of Low Dimensional Quantum Physics and Department of Physics, Tsinghua University, Beijing 100084, China}

\author{Alexei V. Fedorov}
\affiliation{Advanced Light Source, Lawrence Berkeley National Laboratory, Berkeley, CA 94720, USA}

\author{Sung-Kwan Mo}
\affiliation{Advanced Light Source, Lawrence Berkeley National Laboratory, Berkeley, CA 94720, USA}

\author{Eike F. Schwier}
\affiliation{Hiroshima Synchrotron Radiation Center, Hiroshima University, Higashi-Hiroshima 739-0046, Japan}

\author{Mingtian Zheng}
\affiliation{Hiroshima Synchrotron Radiation Center, Hiroshima University, Higashi-Hiroshima 739-0046, Japan}

\author{Yohei Kojima}
\affiliation{Hiroshima Synchrotron Radiation Center, Hiroshima University, Higashi-Hiroshima 739-0046, Japan}

\author{Hideaki Iwasawa}
\affiliation{Hiroshima Synchrotron Radiation Center, Hiroshima University, Higashi-Hiroshima 739-0046, Japan}

\author{Kenya Shimada}
\affiliation{Hiroshima Synchrotron Radiation Center, Hiroshima University, Higashi-Hiroshima 739-0046, Japan}

\author{Kaili Jiang}
\affiliation{State Key Laboratory of Low Dimensional Quantum Physics and Department of Physics, Tsinghua University, Beijing 100084, China}
\affiliation{Collaborative Innovation Center of Quantum Matter, Beijing, P.R. China}

\author{Pu Yu}
\affiliation{State Key Laboratory of Low Dimensional Quantum Physics and Department of Physics, Tsinghua University, Beijing 100084, China}
\affiliation{Collaborative Innovation Center of Quantum Matter, Beijing, P.R. China}

\author{Jia Li}
\affiliation{Key Laboratory of Thermal Management Engineering and Materials, Graduate School at Shenzhen, Tsinghua University, Shenzhen 518055, P.R. China}

\author{Shuyun Zhou*}
\affiliation{State Key Laboratory of Low Dimensional Quantum Physics and Department of Physics, Tsinghua University, Beijing 100084, China}
\affiliation{Collaborative Innovation Center of Quantum Matter, Beijing, P.R. China}

\date{\today}

\begin{abstract}
Epitaxial growth of graphene on transition metal substrates is an important route for obtaining large scale graphene. However, the interaction between graphene and the substrate often leads to multiple orientations, distorted graphene band structure, large doping and strong electron-phonon coupling. Here we report the growth of monolayer graphene with high crystalline quality on Pt(111) substrate by using a very low concentration of an internal carbon source with high annealing temperature. The controlled growth leads to electronically decoupled graphene: it is nearly charge neutral and has extremely weak electron-phonon coupling (coupling strength $\lambda$ $\approx$ 0.056) as revealed by angle-resolved photoemission spectroscopic measurements. The thermodynamics and kinetics of the carbon diffusion process is investigated by DFT calculation. Such graphene with negligible graphene-substrate interaction provides an important platform for fundamental research as well as device applications when combined with a nondestructive sample transfer technique.
\end{abstract}
\maketitle

\section{Introduction}
Graphene, a single layer of carbon atoms arranged in a honeycomb lattice, has attracted extensive research interests due to its intriguing physical properties as well as potential applications \cite{GeimNatureMater07}. Finding a reliable method to produce graphene with large scale and high crystalline quality is one of the central questions for realizing its potential applications \cite{KimPatternGrowth, Seyller, GonGeTransfer}. Epitaxial growth of graphene on transition metals has been considered as a promising route for synthesizing large scale single crystal graphene \cite{ GonMetalSurSci09, GonMetalReview, MullerEpiGraphene}. However, the different lattice constants or orientations often lead to corrugations, ripples, Moir\'{e} patterns, and superlattice bands \cite{ GonMetalSurSci09, GonMetalReview, MullerEpiGraphene}. Moreover, electrons from the metal substrate can interact with $\pi$ electrons in graphene, resulting in charge transfer, strong electron-phonon scattering, band hybridization, and, in some cases, even the absence of Dirac cone \cite{ GonMetalSurSci09, GonMetalReview, MullerEpiGraphene}. Even on the most weakly interacting substrate like Ir(111), clear distortion of the graphene dispersion has also been reported  \cite{PletikosicPRL}. The interaction between graphene and the metal substrate is therefore a major obstacle that needs to be overcome for investigating the fundamental physics of pristine graphene and for realizing its potential applications.

Among all transition metals, Pt(111) is one of the promising substrates for growing quasi-freestanding graphene, since graphene on Pt(111) is expected to have a much larger distance from the substrate compared to other substrates \cite{GonMetalSurSci09, MullerEpiGraphene}. So far graphene on Pt(111)  has been grown mostly  by decomposing or dissolving hydrocarbon molecules \cite{MartenssonPRB08,  SutterPRB09, GaoAPL, GTransfer, GaoGPt, ShikinPRB14}. These methods involve introducing a large amount (up to 0.05$\%$ \cite{GaoGPt}) of an external carbon source, and the dense nucleation sites often lead to graphene with multiple orientations and complicated Moir\'{e} superlattices \cite{SutterPRB09, GaoAPL}, suggesting significant graphene-substrate interaction. In addition, the $\pi$ bands near E$_F$ are obscured by the large intensity contribution from the platinum bands \cite{SutterPRB09}. Summarizing all extensive research mentioned above, we find that so far epitaxial graphene on a metal substrate both with high crystal order and without Moire pattern, charge doping, band hybridization has not been demonstrated yet. Here we report the successful growth of high quality graphene on Pt(111) substrate by utilizing only a very small concentration ($<10^{-5}$ or 10 ppm) of carbon impurities from the high purity (5N) bulk Pt(111) crystal, without introducing any external carbon source. The thermodynamics and kinetics for the segregation process of carbon atoms from the bulk to the surface are simulated by density functional theory (DFT) calculation. This growth method leads to high quality graphene with one dominant orientation rotated by 30$^\circ$ with respect to the Pt(111) substrate, with greatly improved structural and electronic properties compared to previous growth methods. Combining various techniques including low energy electron diffraction (LEED), X-ray photoemission spectroscopy (XPS), Raman spectroscopy, atomic force microscopy (AFM), and angle-resolved photoemission spectroscopy (ARPES), we show that the as-grown graphene is mostly monolayer thick, high quality, nearly charge neutral, and behaves electronically like freestanding graphene with extremely weak electron-phonon coupling. 

\section{Experiment and Calculations}
The epitaxial graphene sample was grown by annealing the Pt(111) substrate in ultra-high vacuum at elevated temperatures up to 1600\textcelsius~using electron beam bombardment. ARPES experiments were performed at our home laboratory with a UV lamp, Beamline 1 of Hiroshima Synchrotron Radiation Center, and Beamlines 12 and 10 of the Advanced Light Source with an energy resolution of 15 meV. The vacuum was maintained below $5\times10^{-11}$ Torr during ARPES measurements and the measurement temperature was 20 K. DFT calculations were performed using Vienna {\it Ab initio} Simulation Package (VASP)\cite{VASPref}. The exchange-correlation potential was treated in the generalized gradient approximation (GGA) of the Perdew-Burke-Ernzerhof(PBE) functional \cite{GGA}. The energy cutoff of the plane-wave expansion was set to 400 eV. The Monkhorst-Pack k-point mesh of 4 $\times$ 4 $\times$ 1 is found to provide sufficient accuracy in the Brillouin zone integration. The climbing image nudged elastic band (CI-NEB) method was used to determine the energy barriers of kinetic processes of carbon atoms escaping from the bulk crystal to the surface \cite{NEB}.

\begin{figure}
\includegraphics[width=8.5 cm] {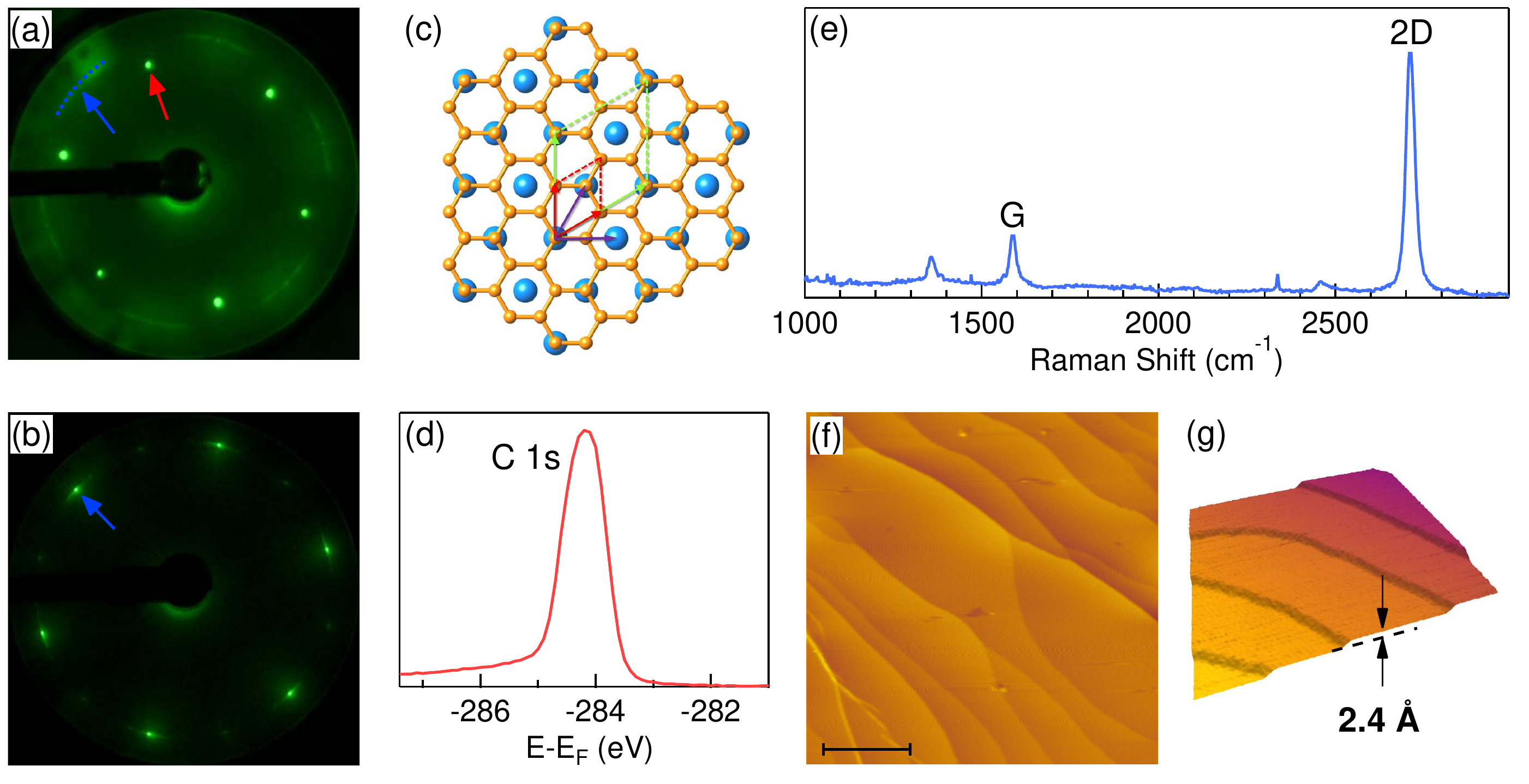}
\caption{ (a) and (b) LEED patterns after annealing Pt(111) crystal to 850 \textcelsius~ (a) and 1600 \textcelsius~ (b) with electron energy of 140 eV.  The red arrow points to the Pt diffraction spot, and blue arrow points to graphene diffraction arc or spot. (c) Schematic Drawing of the crystal structure with superlattice of 2$\times $2/graphene on $\sqrt3\times\sqrt3$R30\textdegree/Pt(111). Blue balls are platinum atoms and orange balls are carbon atoms. The red and green arrows are the unit vectors of the primary cell for graphene and the superlattice. The purple arrows are unit vectors for Pt(111). (d) XPS spectrum of the as-grown graphene measured at a photon energy of 360 eV. (e) Raman spectrum of the as-grown graphene. (f) AFM morphology of the as-grown graphene. The scale bar is 500 nm. (g) Zoom-in 3D view of the AFM image. The step height is about 2.4 \AA~ and the width of the terrace is approximately 300 nm.}
\end{figure}

\section{Results and discussion}
Figure 1 shows the characterization of the sample during and after the growth process. At 850 \textcelsius, a weak arc-shaped diffraction pattern with a larger radius than Pt(111) diffraction pattern emerges in the LEED pattern and coexists with the Pt(111) diffraction spots [Fig.~1(a)]. Using Pt(111) as a reference, the radius of the arcs is calculated to be equal to the reciprocal lattice constant of graphene 2.94 \AA$^{-1}$, suggesting that the arc-shaped pattern may arise from graphene-like patches on the Pt(111) surface. The arcs instead of discrete spots suggest that the graphene patches have multiple domains with different azimuthal orientations, which is similar to previous studies using an external carbon source \cite{SutterPRB09, GaoAPL}. After further annealing at a much higher temperature of 1600  \textcelsius, strong graphene diffraction spots emerge [Fig.~1(b)], suggesting that graphene domains aggregate and high azimuthal order is developed. The dominant set of diffraction spots (marked by blue arrow in Fig.~1(b)) are rotated by 30\textdegree~from the Pt(111) diffraction spots. Traces of another set of diffraction spots, which are at 0\textdegree~from the Pt(111) orientation, are also observed but with a much weaker intensity, and its relative intensity to the R30\textdegree~domain can be minimized by optimizing the growth conditions. The structure of the R30\textdegree~domain is in agreement with 2$\times$2/graphene on $\sqrt3\times\sqrt3$R30\textdegree/Pt [Fig.~1(c)], which was reported to have the weakest corrugation compared to other graphene structures by scanning tunneling microscopy (STM) measurements \cite{GaoAPL}. 

The successful growth of graphene is further confirmed by XPS and Raman measurements. The XPS in Fig.~1(d) shows a strong carbon 1s core level peak at binding energy of a 284 eV. The Raman spectrum in Fig.~1(e) shows characteristic G and 2D peaks of graphene \cite{FerrariRaman}. The line-shape and the width of the 2D peak are directly related to the thickness of the graphene sample \cite{FerrariRaman, GrafRaman} -  for monolayer graphene, the 2D peak is symmetric and the width is $\approx$ 30 cm$^{-1}$, while for bilayer graphene, the 2D peak is asymmetric and the width is  $\approx$ 60 cm$^{-1}$ \cite{FerrariRaman, GrafRaman}. The symmetric and sharp 2D peak with average full width half maximum of 35 cm$^{-1}$ [Fig.~1(e)] confirms that our sample is mostly monolayer thick. Figure 1f shows the surface morphology of the as-grown graphene sample using AFM. Flat terraces with size of a few hundred nanometers are clearly observed, indicating the high quality of the graphene sample. The step height of 2.4 \AA~ is very close to one layer thickness of platinum (2.26 \AA), and is very different from the distance between graphene and Pt(111) substrate (3.7 \AA)  \cite{MullerEpiGraphene}, or the separation between graphene layers (3.45 \AA). This suggests that the surface is almost entirely covered by uniform monolayer graphene, which is also consistent with the absence of Pt(111) diffraction spots in the LEED pattern.

\begin{figure}
\includegraphics[width=8.0 cm] {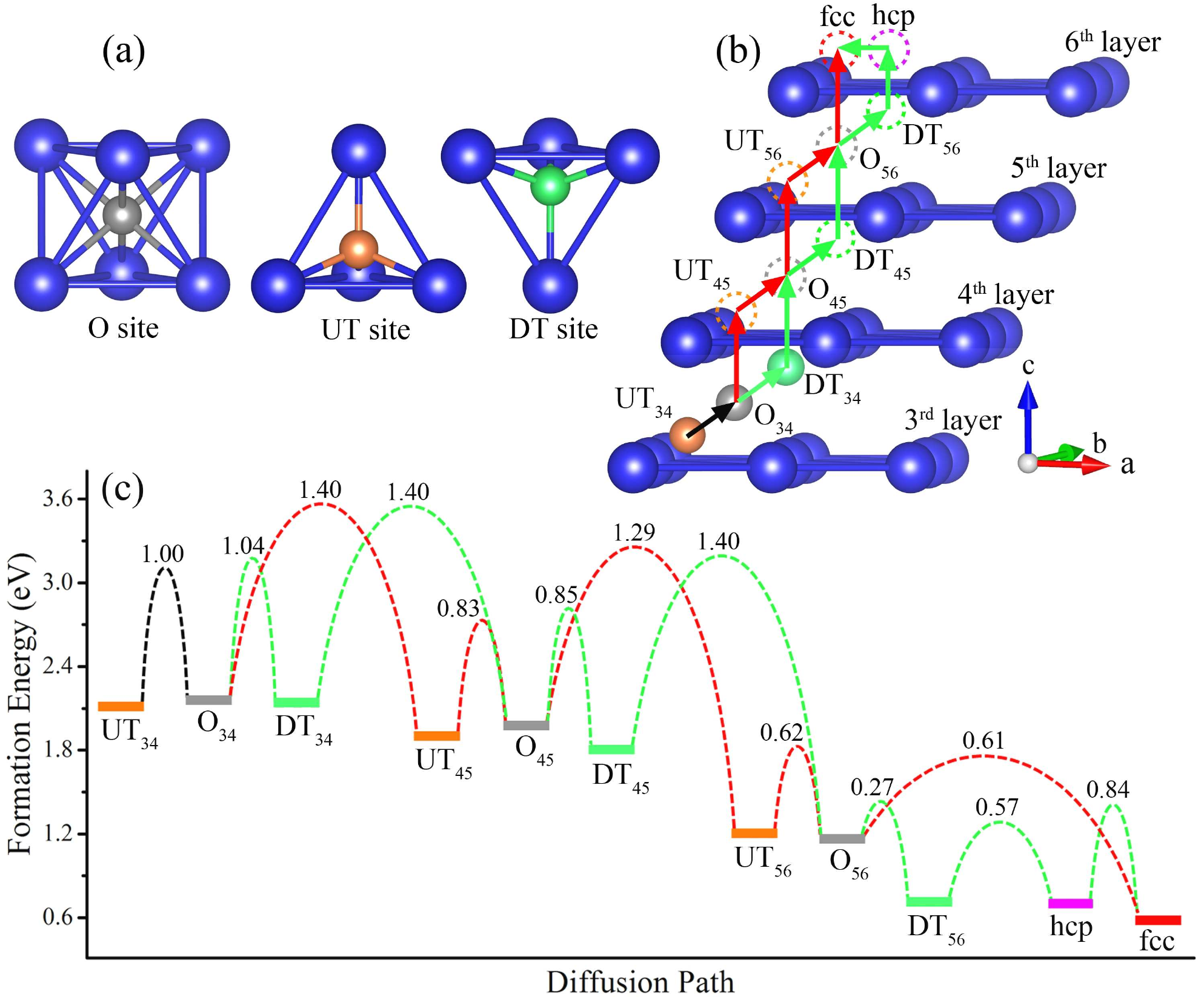}
\caption{ (a) Schematic drawings of the 3 possible sites for carbon atom located within the interlayer space of Pt(111) --- Octahedral site (O) and two tetrahedral sites (UT and DT) . The blue ball is for platinum atom and the orange, gray and green balls are for carbon atoms. (b) Possible diffusion paths for carbon atoms escaping from the UT$_{34}$ site to the surface. (c) Formation energy of carbon atoms and energy barriers between carbon sites for intralayer and interlayer diffusion. }
\end{figure}

In order to reveal the initial process of the graphene growth, we thus construct a Pt(111) six-layer slab model to calculate the formation energies and energy barriers of isolated carbon atoms escaping from the bulk crystal to the surface using DFT calculation. As shown in Fig.~2(a), there are three high symmetry sites for carbon atoms in the interlayer space of Pt(111): Octahedral (O), upward tetrahedral (UT) and downward tetrahedral (DT)) sites. The formation energies of carbon atoms located at the O$_{34}$, UT$_{34}$ and DT$_{34}$ (octahedral site and tetrahedral sites between the third and fourth layers) are 2.16 eV, 2.13 eV, and 2.11 eV, respectively. These are almost identical to those of  the carbon atoms located in the bulk Pt crystal, 2.18 eV (octahedral site) and 2.20 eV (tetrahedral site), suggesting that the third and fourth layers in the six-layer slab are sufficient to represent the bulk properties. As shown in Fig.~2(c), the value of the formation energy decreases when the carbon atoms escape from the bulk to the surface, suggesting that carbon atoms thermodynamically prefer to locate on the surface, instead of in the bulk. The most stable site for isolated carbon atoms is the fcc hollow (fcc) site on the surface \cite{Angew2011}, which can act as nucleation sites during the graphene growth process. To reveal the kinetic process, we have also calculated the energy barriers of all possible diffusion paths, including surface diffusion, the diffusion within the layer, and the diffusion across the Pt layer. The two most possible diffusion paths which have the lowest energies are identified in Fig.~2(b). The energy barriers are shown as broken lines in Fig.~2(c), and the maximum barrier for interlayer carbon diffusion is 1.40 eV, which is larger than the maximum energy barrier of 1.04 eV for intralayer diffusion. Thus, the percentage of possible carbon atoms escaping from the bulk crystal to the surface in all diffusing carbon atoms can be approximately estimated by $\exp (-{\triangle}E_{\rm B}/k_{\rm B}T){\approx} 8\%$, where $T$ is taken as our typical experimental temperature of 1600 \textcelsius. Our calculation suggests that carbon impurities in Pt crystal can diffuse from the bulk to the surface, analyzed from both thermodynamic and kinetic viewpoints, to form the graphene layer.

\begin{figure}
\includegraphics[width=8.5 cm] {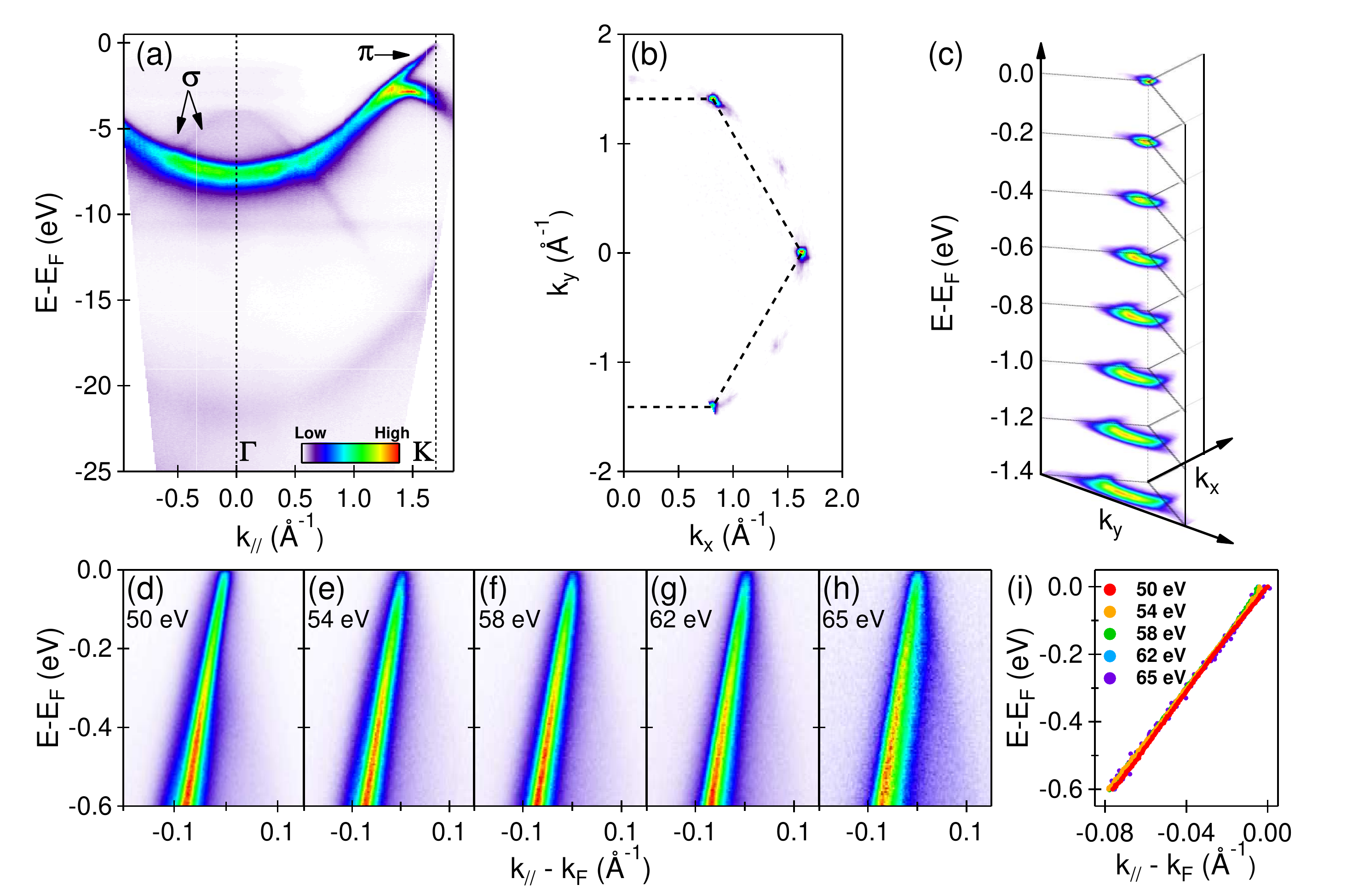}
\caption{ (a) Band structure of the graphene on Pt(111) along the $\Gamma$-$K$ direction. The clear $\pi$ band and $\sigma$ band are indicated by the arrows. The other $\pi$ band at high binding energy comes from the R0\textdegree~domain along its $\Gamma$-$M$ direction. (b) Measured point-like Fermi surface map of the graphene. The dashed line indicates the Brillouin zone boundary of the R30\textdegree~domain. (c) Conical dispersion near the $K$ point. (d)-(h) ARPES data measured along the $\Gamma$-$K$ direction at photon energies of 50 eV, 54 eV, 58 eV, 62 eV, and 65 eV respectively. The corresponding reduced $k_z$ values are 0.9 c$^*$, 0.05 c$^*$,  0.20 c$^*$, 0.34 c$^*$ and 0.44 c$^*$ (c$^*$= 2$\pi$/6.708\AA{} = 0.937 \AA$^{-1}$). (i) Extracted dispersions from data shown in (d)-(h).}
\end{figure}

The nearly ideal electronic structure of the as-grown graphene is further revealed by ARPES measurement. Figure 3a shows the band structure of graphene  measured along $\Gamma$-$K$ direction in a wide energy range. The characteristic $\pi$ bands and $ \sigma$ bands of graphene are observed clearly. Different from graphene grown on other metal substrates \cite{HofmannPRB13, EliPRB}, our data show negligible contribution from the Pt(111) substrate bands or Moir\'{e} superlattice bands, which makes it more convenient to probe the electronic structure of graphene. Peaks from $\pi$ band along the $\Gamma$-$M$ direction of the R0\textdegree~domain are also observed at higher binding energy. In addition, there are weak and non-dispersive peaks at  -10.7, -3.0, and -1.7 eV, which are likely caused by impurity scattering. Figure 3(b) shows the measured Fermi surface of graphene. The observation of a stronger set of Dirac cones from the R30\textdegree~domain and a weaker one from the R0\textdegree~domain is consistent with LEED pattern. The point-like Fermi surface shows that  the graphene is almost charge neutral and there is negligible charge transfer from the substrate. Figure 3(c) shows the conical dispersion at the $K$ point. No splitting of the cones is observed, suggesting that the graphene sample is monolayer thick. The monolayer thickness of the graphene sample is further verified by the absence of $k_z$ dependence shown in Figs.~3(d)-3(i).  No splitting of the $\pi$ bands is observed for all photon energies which cover a $k_z$ range from the K point to the H point in graphite Brillouin Zone [Figs.~3(d)-3(h)], and the extracted dispersions [Fig.~3(i)] overlap with each other, confirming that the majority of the graphene sample is monolayer, which is in agreement with Raman and AFM measurements.

\begin{figure}
\includegraphics[width=8.8 cm] {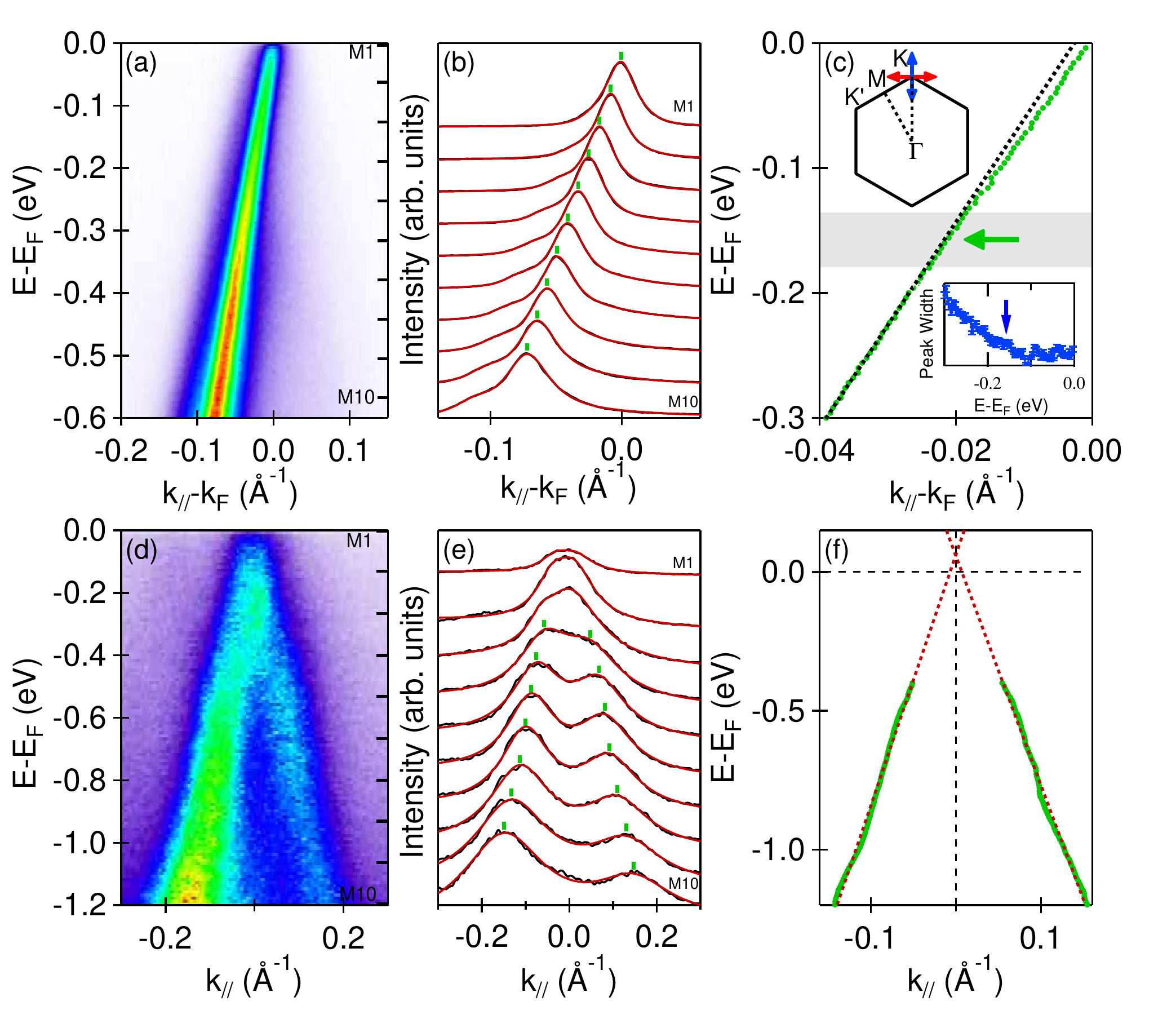}
\caption{(a) ARPES data along the $\Gamma$-$K$ direction. (b) MDCs (black) and the fitting curves (red). The green markers show the peak positions. (c) Extracted dispersion. The upper inset shows the cut direction of (a) and (d). The lower inset shows the MDC peak width of (b). (d) Linear dispersion near $K$ point (vertical to $\Gamma$-$K$ direction). (e) MDCs (black) and the fitting curves (red). (f) Extracted dispersion. The Dirac point locates at 60 ($\pm$5) meV above Fermi level. }
\end{figure}

Figure 4 shows analysis of the electronic structure near the Dirac cone along two high symmetric directions. Figure 4(a) shows ARPES data measured along the $\Gamma$-$K$ direction. Figure 4(b) shows the momentum distribution curves (MDCs), which can be fitted by a Lorentzian peak from the monolayer graphene and a much weaker one on the left which likely comes from a small amount of bilayer graphene at the edges. From the extracted dispersion [Fig.~4(c)], the Fermi velocity is extracted to be 1$ \times $10$^{6}$ m/s, which is very close to that of pristine graphene. Moreover, a sudden change of Fermi velocity and a decreased scattering rate (see peak width in Fig.~4(c)) at a binding energy around 160 meV [Fig.~4(c)] are observed, which are characteristic features of electron-phonon interaction in graphene \cite{SZhouKohn, LanzaraElphPNAS, LiuFPRL2013}. From the velocity renormalization, the electron-phonon coupling strength is extracted to be very weak ($\lambda$ $\sim$ 0.056) compared with other epitaxial graphene ($\lambda$ ranging from 0.14 to 0.3) \cite{EliNatPhys,SZhouKohn,IrkinkPletikosic,IrkinkHofmann}. We note that the strength of electron-phonon coupling increases strongly with carrier concentration \cite{LanzaraElphPNAS, LiuFPRL2013}, and the observation of extremely weak electron-phonon interaction is in agreement with the small doping of the quasi-freestanding graphene. The low doping level is further confirmed by data taken perpendicular to $\Gamma$-$K$ direction, where dispersions from both sides of the cone can be clearly resolved [Fig.~4(d)]. By extrapolating the extracted dispersions, the Dirac point energy is found to be at 60 meV above the Fermi level [Fig.~4(f)]. This is much closer to the Fermi level than the 300 meV reported in graphene/Pt(111) previously \cite{SutterPRB09}. Such almost charge-neutral, electronically decoupled graphene with extremely weak electron-phonon interaction is ideal for investigating the intrinsic properties of graphene. Since electron-phonon interaction has great impact on its transport properties, the reduction of carrier concentration and reduced electron-phonon interaction are also significant for device applications.

Compared to previous graphene samples on Pt(111) which were grown by using a large amount of external carbon source \cite{MartenssonPRB08, GaoAPL, ShikinPRB14, GTransfer, GaoGPt,  SutterPRB09}, the graphene sample grown from the small concentration of internal carbon source shows distinguished properties: negligible interaction with the substrate, one dominant orientation, almost charge neutral and extremeley weak electron-phonon interaction ($\lambda$ $\sim$ 0.056). The controlled growth of electronically decoupled graphene is achieved by using two critical growth conditions. First, a much smaller carbon concentration ($< 10^{-5} $) from the bulk (instead of externally induced large carbon concentration) and thus sparse nucleation sites during the growth process. Second, high annealing temperature (1600 \textcelsius) compared to that used in previous studies (\textless 1000 \textcelsius), which makes the graphene highly oriented. We note that similar segregation of carbon impurities from the bulk to the surface has been applied for growing graphene on Ru(0001) and Ir(111) \cite{STMGRu,STMGIr}. However, the large corrugation, multiple orientations of graphene and stronger electron-phonon interaction on those substrates undermine the significance of this growth method.  We believe that these conditions above can improve the growth of graphene on other transition metal substrates. 

\section{summary}
To summarize, we report the growth of monolayer graphene on platinum substrate with nearly ideal graphene band structure. Such graphene is important for both fundamental research and applications. First, it provides a platform not only for investigating the properties of pristine graphene, including many-body interaction, Dirac-fermion physics, but also for constructing a variety of quasi-freestanding Van der Waals heterostructures with other 2D materials \cite{Geimvanderwaals, NLvanderwaals}. Second, by combining well-developed graphene transfer methods \cite{GTransfer, facetofaceTransfer, Drytransfer}, for example, the bubbling transfer based on a water electrolysis process which is nondestructive to both graphene and the Pt(111) substrate \cite{GTransfer}, the as-grown epitaxial graphene can be transferred to other substrate for device applications, and the Pt(111) substrate can be recycled for repeated growth.
\\

\section*{Acknowledgments}
This work is supported by the National Natural Science Foundation of China (11274191, 11334006 and 11104155), Ministry of Education of China (20121087903, 20121778394) Ministry of Science and Technology of China (2011CB606405), and Shenzhen Projects for Basic Research (JCYJ20120831165730910 and KQCX20140521161756227). The Advanced Light Source is supported by the Director, Office of Science, Office of Basic Energy Sciences, of the US Department of Energy under Contract No. DE-AC02-05CH11231. Experiments at HiSOR were performed through proposals No. 13-B-20 and No. 14-A-10.  E.F.S. acknowledges financial support by the JSPS postdoctoral fellowship for overseas researchers as well as the Alexander von Humboldt Foundation. 

{\bf Corresponding Author}
\\$ \ast $E-mail: syzhou@mail.tsinghua.edu.cn

\bibliography{Ref}
\bibliographystyle{apsrev4-1}

\end{document}